# $Ba_6RE_2Ti_4O_{17}$ (RE= Nd, Sm, Gd, Dy-Yb): A family of Rare-Earth-based layered triangular lattice magnets

Fangyuan Song,[†] Andi Liu,[†,‡] Qiao Chen, [§] Jin Zhou,[†] Jingxin Li,[#] Wei Tong,[#] Shun Wang,[§] Yanhong, Wang,[‖] Hongcheng, Lu,[‖] Songliu Yuan,[†] Hanjie Guo,[‡*] Zhaoming Tian[†*]

[†] School of Physics and Wuhan National High Magnetic Field Center, Huazhong University of Science and Technology, Wuhan, 430074, China.

[‡] Songshan Lake Materials Laboratory, Dongguan, Guangdong 523808, China

[§] School of Physics and MOE Key Laboratory of Fundamental Physical quantum Physics, PGMF, Huazhong University of Science and Technology, Wuhan, 430074, China.

[#] Anhui Province Key Laboratory of Condensed Matter Physics at Extreme Conditions, High Magnetic Field Laboratory, Chinese Academy of Sciences, Hefei, 230031, China.

[‖]Key Laboratory of Material Chemistry for Energy Conversion and Storage, School of Chemistry and Chemical Engineering, Huazhong University of Science and Technology, Wuhan, 430074, China.

**ABSTRACT:** The exploration of new rare-earth (RE)-based triangular-lattice materials plays a significant role in motivating the discovery of exotic magnetic states. Herein, we report a family of hexagonal perovskite compounds $Ba_6RE_2Ti_4O_{17}$ (RE = Nd, Sm, Gd, Dy-Yb) with space group of $P6_3$/mmc, where magnetic $RE^{3+}$ ions are distributed on the parallel triangular-lattice layers within the *ab*-plane and stacked in an "AA"-type fashion along the *c*-axis. The low-temperature magnetic characterizations indicate that all synthesized $Ba_6RE_2Ti_4O_{17}$ compounds exhibit the dominant antiferromagnetic (AFM) interactions and the absence of magnetic order down to 1.8 K. The isothermal magnetization and electron spin resonance results reveal the distinct magnetic anisotropy for the compounds with different RE ions. Moreover, the as-grown $Ba_6Nd_2Ti_4O_{17}$ single crystals exhibit the Ising-like magnetic anisotropy with magnetic easy-axis perpendicular to the triangle-lattice plane and no long-range magnetic order down to 80 mK, as the quantum spin liquid candidate with dominant Ising-type interactions.

## ■ INTRODUCTION

The triangular-lattice antiferromagnets have attracted great attention to search for the exotic magnetic states and emergent magnetic phenomena. One of archetypical case is the quantum spin liquid (QSL) state, which refers to a highly spin-entangled state but without magnetic order down to the lowest temperature. Since the initial theoretical proposal of QSL state in two-dimensional (2D) triangular-lattice Heisenberg antiferromagnets by P. W. Anderson,[1] significant research efforts have been made to realize this exotic state in real materials especially the antiferromagnetically coupled spin-1/2 triangular lattice systems. The typical examples for such states include the organic compounds $\kappa$–(BEDT-TTF)$_2$Cu$_2$(CN)$_3$, EtMe$_2$Sb[Pd(dmit)$_2$]$_2$ and inorganic $NiGa_2S_4$, $Ba_3CuSb_2O_9$ compounds.[2-5] Additionally, the triangular-lattice antiferromagnets can exhibit exotic magnetism including the novel up-up-down (uud) ground state stabilized by quantum fluctuations and field-induced quantum phase transition behaviors, which is manifested by a one-third (1/3) fractional magnetization over a finite magnetic field range as experimentally confirmed in $Cs_2CuBr_4$ and

$Na_2BaCo(PO_4)_2$,[6,7] as so on.

Compared to large numbers of studies on $S$=1/2 triangular-lattice magnets containing transition-metal (TM) ions, in past decade, the rare earth (RE)-based materials have also attracted a growing attention to explore the novel magnetic states. The large spin-orbit coupling and crystal electric field (CEF) effects of 4f electrons of RE ions can lead to highly anisotropic exchange interactions. For the compounds including Kramers ions with odd 4*f* electrons (like $Ce^{3+}$, $Nd^{3+}$, and $Yb^{3+}$), its low temperature magnetism can be well described by a spin−orbit coupled $J_{eff}$ = 1/2 moments, thus provide the platform to realize the novel magnetic states beyond the 3d TM-based ones. Until now, several RE-based magnets with triangle lattice have been studied to search for the QSL states, such as $YbMgGaO_4$,[8] $AYbCh_2$ (A = $Li^+$, $Na^+$, $K^+$, $Cs^+$, Ch = O, S, Se;)[9-12] and $PrZnAl_{11}O_{19}$,[13] etc. Moreover, some of compounds are interesting due to the observation of novel quantum magnetic phases like the Berezinskii-Kosterlitz-Thousless (BKT) phase in $TmMgGaO_4$ and spin glass phase in $YbZnGaO_4$ and quantum dipolar liquid in $Yb(BaBO_3)_3$ systems,[14-16] etc. Meanwhile, the experimental identifications on the above exotic magnetic states remain to be controversial due to the unavoidable intrinsic antisite disorder, structural distortion and complex superexchange pathway in real materials.[17] Based on this consideration, new scenarios on materials' design are required to explore the novel magnetic phases. In this respect, very recently, one route on realizing the QSL state based on the dominant Ising-type correlations instead of Heisenberg-type interactions has been experimentally investigated on triangular-lattice $NdTa_7O_{19}$,[18] which is instructive to search for the novel magnetic states but limited materials have been identified. From the structure viewpoint, the stacking order of triangular-lattice multilayers as well as the formation of bilayer structure have been proposed to introduce the magnetic frustration and generate diverse exotic magnetic states.[19,20] Therefore, designing new RE-triangular-lattice materials with different stacking structures and magnetic anisotropy are highly desirable to enrich the exotic magnetic phenomena.

In this work, we report a family of RE-based triangular lattice compounds $Ba_6RE_2Ti_4O_{17}$ (RE = Nd, Sm, Gd, Dy-Yb) crystallized into the hexagonal crystal structure (space group $P6_3/mmc$). Among the family, two members $Ba_6RE_2Ti_4O_{17}$ (RE=Nd, Y) have been synthesized and structurally characterized to be the 12-layer hexagonal perovskite structure by X-ray and neutron powder diffractions.[21] While, the lattice geometry of magnetic $RE^{3+}$ ions and their magnetism have not been unveiled, motivating the present study. Here, on the basis of structure analysis, we reveal that magnetic $RE^{3+}$ ions form the parallel triangular-lattice layers with "AA"-type stacking fashion along the *c*-axis and the neighboring interlayer magnetic planes are separated alternatively by the nonmagnetic $Ti_2O_9$ dimers and double sheets of $TiO_4$ tetrahedra. For all $Ba_6RE_2Ti_4O_{17}$ compounds, magnetic measurements and electron spin resonance (ESR) spectra indicate the dominant antiferromagnetic (AFM) exchange interactions and absence of magnetic order down to 1.8 K. Moreover, the $Ba_6Nd_2Ti_4O_{17}$ single crystals have been successfully grown and which exhibit the Ising-like anisotropy with easy magnetization along the *c*-axis and no long-range magnetic order down to 0.08 K.

## ■ EXPERIMENTAL SECTION

**Material Synthesis.** The series of $Ba_6RE_2Ti_4O_{17}$ (RE = Nd, Sm-Gd, Dy-Yb) polycrystals were synthesized by the conventional solid-state reaction using the $BaCO_3$ (99.9%), $TiO_2$ (99.9%) and RE oxides (RE=Nd, Sm- Gd,

Dy-Yb, 99.9%) as starting materials. No uncommon hazards are noted in the experiment on sample synthesis.The RE oxides were dried at 800 °C overnight prior to use. The stoichiometric mixtures of the raw materials were thoroughly ground for two hours, pressed into pellets and preheated at 1100°C for 24 h. For $Ba_6RE_2Ti_4O_{17}$ (RE = Nd, Sm-Gd) samples, the products were reground and calcined in air at temperature regions from 1250 to 1300°C for 4 days. For RE = Dy-Yb samples, the higher reacting temperatures of 1350 -1400 °C are required to obtain the pure phase sample. All the synthesized $Ba_6RE_2Ti_4O_{17}$ samples are stable in air, and the optical images of the $Ba_6RE_2Ti_4O_{17}$ pellets are shown in Figure S1 (see supporting information).

The $Ba_6Nd_2Ti_4O_{17}$ single crystal was grown by high-temperature flux method using $BaCl_2$ as flux, similar to the previous descriptions.[22] The mixtures of $Ba_6Nd_2Ti_4O_{17}$ polycrystaline powders and $BaCl_2$ flux with a ratio of 1:15 were placed in a 50-ml platinum crucible and then heated to1250 °C for 24 hours to ensure the homogeneous melting. Then, the temperature of furnace was cooled slowly to 800 °C at the rate of 3 °C /h, and finally cooled down to room temperature. The $Ba_6Nd_2Ti_4O_{17}$ single crystals were separated by dissolving the $BaCl_2$ flux in hot water. The as-grown $Ba_6Nd_2Ti_4O_{17}$ single crystal has the hexagonal shape, as shown in the inset of Figure 1a.

**Structure Characterization.** The crystal structure and phase purity of $Ba_6RE_2Ti_4O_{17}$ (RE = Nd, Sm, Eu, Gd, Dy-Yb) compounds were analyzed by Powder X-ray diffraction (PXRD, SmartLab) data at room temperature with Cu$K_\alpha$ ($\lambda$ = 1.5418 Å). The structural refinements for $Ba_6RE_2Ti_4O_{17}$ were carried out by Rietveld method with the GSAS program. The diffraction data of $Ba_6Nd_2Ti_4O_{17}$ single crystals were examined by single-crystal X-ray diffraction (SXRD) using a Bruker SMART APEX DUO diffractometer equipped with a CCD detector (graphite-monochromated Mo$K_\alpha$ radiation, $\lambda$ = 0.71073 Å). The structure was refined by the ShelXL least squares software package with the Olex2 program.

**Physical Property Measurements.** Magnetic properties were measured with a commercial superconducting quantum interference device (SQUID, Quantum Design) magnetometer. The dc magnetic susceptibilities $\chi(T)$ of $Ba_6RE_2Ti_4O_{17}$ polycrystals were measured under applied field of 1 kOe in temperature range from 1.8 to 300 K. The isothermal magnetization $M(\mu_0H)$ measurements were collected by SQUID with field up to 7 T. Low-temperature X-band electron paramagnetic resonance (EPR) spectra at a microwave frequency of 9.4 GHz were carried out using a Bruker spectrometer equipped with an Oxford ESR910 liquid helium cryostat. The specific heat $C_p(T)$ measurements of $Ba_6Nd_2Ti_4O_{17}$ compounds were performed using a Quantum Design physical properties measurement system (PPMS) in the temperature range 2 - 150 K at different magnetic fields. At low temperatures down to 0.08 K, the specific heat was measured using the PPMS equipped with a dilution refrigeration by the heat capacity option.

## ■ RESULTS AND DISCUSSION

**Structure Description.** The synthesized $Ba_6RE_2Ti_4O_{17}$ (RE = Nd, Sm, Gd, Dy-Yb) polycrystalline samples are isostructural and crystalized into the hexagonal crystal structure with space group $P6_3/mmc$ (No.194). Figure 1a shows the experimental and the refined XRD patterns for the selected $Ba_6RE_2Ti_4O_{17}$ (RE = Nd, Sm, Ho) samples. Here, the crystal structure of $Ba_6Y_2Ti_4O_{17}$ was adopted as an initial model for Rietveld refinement,[21] the refinements yield a good fit to the observed experimental data. The obtained lattice parameters and atomic positions of $Ba_6RE_2Ti_4O_{17}$ samples are listed in Table S1 (see supporting information). As reduced $RE^{3+}$ ionic radii, both the lattice constants (*a*, *c*) and unit-cell volume (*V*) follow a monotonically decrease (see Figure 1b, c). The selected bond distances, bond angles and intralayer and interlayer RE-RE distances for $Ba_6RE_2Ti_4O_{17}$ polycrystals are summarized in Table 1. For the grown $Ba_6Nd_2Ti_4O_{17}$ single crystals, the refined crystallographic data is presented in Table 2.

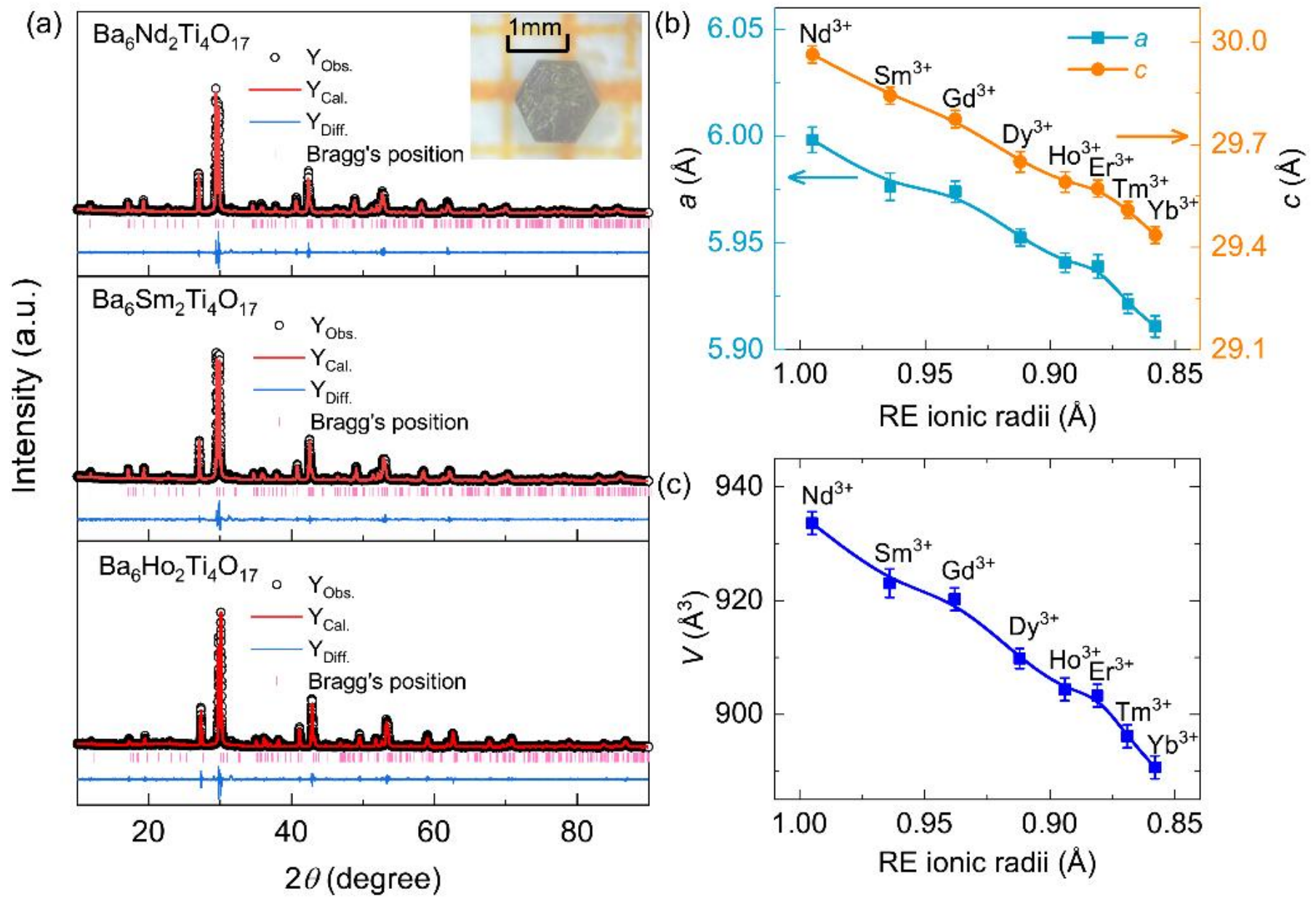

**Figure 1**. (a) The experimental and refined powder X-ray diffraction (XRD) patterns of $Ba_6RE_2Ti_4O_{17}$ (RE = Nd, Sm, Ho) compounds, the experimental data is shown by black open circles, the refined data is in red line and the difference in blue line. Inset shows the photograph of as-grown $Ba_6Nd_2Ti_4O_{17}$ single crystal on 1 mm graph paper. (b) The variation of lattice constants and (c) the unit-cell volume versus RE ionic radii of $Ba_6RE_2Ti_4O_{17}$.

Table 1. Some bond distances, bond angles, and RE-RE distances of polycrystalline $Ba_6RE_2Ti_4O_{17}$ (RE = Nd, Sm, Gd, Dy-Yb).

| RE | Nd | Sm | Gd | Dy | Ho | Er | Tm | Yb |
|---|---|---|---|---|---|---|---|---|
| $REO_6$ | | | | | | | | |
| RE-$O_2$×3 | 2.377(5) | 2.371(2) | 2.296(2) | 2.341(1) | 2.231(9) | 2.263(3) | 2.375(2) | 2.232(7) |
| RE-$O_3$×3 | 2.325(0) | 2.259(5) | 2.244(2) | 2.142(3) | 2.224(6) | 2.167(1) | 2.175(5) | 2.244(5) |
| Average RE-O | 2.351(2) | 2.315(3) | 2.270(2) | 2.241(7) | 2.228(3) | 2.215(2) | 2.275(4) | 2.238(6) |
| Interplane RE−RE (Å) | 7.547(2) | 7.411(2) | 7.370(3) | 7.352(2) | 7.323(0) | 7.319(5) | 7.307(5) | 7.301(7) |
| | 7.434(5) | 7.510(3) | 7.516(8) | 7.449(5) | 7.472(1) | 7.466(2) | 7.446(8) | 7.416(0) |
| intraplane RE−RE (Å) | 5.998(2) | 5.976(2) | 5.974(0) | 5.952(5) | 5.940(7) | 5.939(0) | 5.921(6) | 5.910(9) |
| $O_2$–RE–$O_3$ (deg) | 95.5(1) | 94.4(2) | 94.6(5) | 89.3(8) | 94.7(7) | 94.7(7) | 93.2(1) | 95.9(5) |
| Ti(1)$O_4$ | | | | | | | | |
| Ti(1)-$O_1$×1 | 1.654(9) | 1.720(4) | 1.746(2) | 1.50(4) | 1.866(7) | 1.689(6) | 1.745(5) | 1.806(0) |
| Ti(1)-$O_2$×3 | 1.808(4) | 1.762(5) | 1.835(2) | 1.758(2) | 1.840(2) | 1.813(1) | 1.691(4) | 1.820(7) |
| Average Ti(1)-O | 1.770(0) | 1.751(9) | 1.812(8) | 1.694(6) | 1.846(8) | 1.782(2) | 1.704(9) | 1.817(0) |
| $O_1$–Ti–$O_2$ (deg) | 108.4(8) | 109.6(3) | 108.5(9) | 112.4(8) | 110.7(9) | 109.1(4) | 108.7(0) | 106.2(6) |
| Ti(2)$O_6$ | | | | | | | | |
| Ti(2)-$O_3$×3 | 1.862(6) | 1.928(4) | 1.928(2) | 2.140(3) | 1.936(0) | 1.974(7) | 1.943(8) | 1.873(5) |
| Ti(2)-$O_4$×3 | 1.990(7) | 1.960(0) | 1.985(1) | 1.947(2) | 2.033(2) | 1.995(5) | 1.961(4) | 1.994(8) |
| Average Ti(2)-O | 1.926(7) | 1.944(2) | 1.956(7) | 2.043(8) | 1.984(6) | 1.985(1) | 1.952(6) | 1.934(2) |
| $O_3$–Ti(2)–$O_4$ (deg) | 93.4(4) | 93.6(4) | 92.8(6) | 95.3(8) | 90.5(0) | 91.9(7) | 93.9(9) | 91.4(2) |
| Ba(1)$O_6$ | | | | | | | | |
| Ba(1)-$O_2$×6 | 2.780(3) | 2.836(8) | 2.836(2) | 2.902(2) | 2.866(1) | 2.835(1) | 2.847(6) | 2.775(1) |
| Average Ba(1)-O | 2.780(3) | 2.836(8) | 2.836(2) | 2.902(2) | 2.866(1) | 2.835(1) | 2.847(6) | 2.775(1) |
| $O_2$-Ba(1)-$O_2$ (deg) | 65.9(7) | 66.2(6) | 63.7(7) | 65.5(6) | 62.1(8) | 63.2(1) | 67.0(6) | 62.6(1) |
| Ba(2)$O_{10}$ | | | | | | | | |
| Ba(2)-$O_1$×1 | 2.569(1) | 2.515(7) | 2.498(2) | 2.673(3) | 2.407(9) | 2.594(2) | 2.544(3) | 2.545(2) |
| Ba(2)-$O_2$×6 | 3.036(5) | 3.019(5) | 3.011(9) | 2.994(0) | 2.988(5) | 2.991(7) | 2.996(1) | 2.985(5) |
| Ba(2)-$O_3$×3 | 2.948(1) | 2.934(3) | 2.935(2) | 2.76(4) | 2.959(4) | 2.934(2) | 2.848(5) | 2.912(0) |
| Average Ba(2)-O | 2.963(2) | 2.943(6) | 2.937(5) | 2.892(9) | 2.921(7) | 2.934(7) | 2.906(6) | 2.919(4) |
| $O_1$–Ba(2)–$O_3$ (deg) | 81.0(4) | 81.8(5) | 82.6(4) | 84.0(3) | 83.7(3) | 83.0(0) | 81.4(3) | 81.8(9) |
| $O_1$–Ba(2)–$O_3$ (deg) | 147.7(4) | 146.5(4) | 146.2(4) | 144.8(7) | 145.9(6) | 144.9(1) | 144.9(4) | 146.1(7) |
| $O_2$–Ba(2)–$O_3$ (deg) | 71.1(3) | 69.6(2 | 68.3(5) | 66.3(6) | 66.9(3) | 59.5(6) | 68.9(3) | 68.6(4) |
| Ba(3)$O_9$ | | | | | | | | |
| Ba(3)-$O_3$×6 | 3.030(7) | 3.023(2) | 3.019(4) | 3.060(9) | 2.993(3) | 2.999(0) | 2.993(1) | 2.975(9) |
| Ba(3)-$O_4$×3 | 2.807(5) | 2.794(9) | 2.776(1) | 2.644(2) | 2.710(2) | 2.759(9) | 2.811(8) | 2.767(0) |
| Average Ba(3)-O | 2.926(3) | 2.947(1) | 2.938(3) | 2.922(0) | 2.898(9) | 2.919(3) | 2.932(7) | 2.906(3) |
| $O_3$–Ba(3)–$O_4$ (deg) | 57.3(1) | 58.1(7) | 58.3(5) | 58.8(2) | 59.0(1) | 59.2(6) | 58.8(4) | 57.5(5) |
| Ba(4)$O_{12}$ | | | | | | | | |
| Ba(4)-$O_3$×6 | 3.012(4) | 3.007(6) | 2.985(2) | 3.17(4) | 2.918(2) | 2.941(8) | 2.994(6) | 2.914(6) |
| Ba(4)-$O_4$×6 | 3.012(5) | 3.002(1) | 2.998(6) | 2.979(7) | 2.974(9) | 2.979(1) | 2.975(3) | 2.965(7) |
| Average Ba(4)-O | 3.012(4) | 3.004(9) | 2.991(9) | 3.076(9) | 2.9465(5) | 2.960(4) | 2.984(9) | 2.940(2) |
| $O_3$-Ba(4)-$O_4$ (deg) | 55.3(9) | 56.3(1) | 56.5(4) | 58.8(2) | 57.1(6) | 57.6(5) | 62.0(5) | 56.2(1) |

Table 2. Crystal data and structure refinements of $Ba_6Nd_2Ti_4O_{17}$ single crystal.

| Empirical formula | $Ba_6Nd_2Ti_4O_{17}$ |
|---|---|
| Formula weight | 1576.12 |
| *T* (K) | 296 |
| $\lambda$ (Å) | 0.711073 |
| Space group | $P6_3/mmc$ |
| *a* (Å) | 5.9934(10) |
| *b* (Å) | 5.9934(10) |
| *c* (Å) | 29.952(5) |
| $\alpha$ (°) | 90 |
| $\beta$ (°) | 90 |
| $\gamma$ (°) | 120 |
| $V$ (Å$^3$) | 931.8(3) |
| $Z$ | 2 |
| Density (g cm$^{-3}$) | 5.618 |
| Absorption coefficient (mm$^{-1}$) | 19.596 |
| GOF on $F^2$ | 1.256 |
| $R_{1,}$ | 0.0814 |
| $wR_2$ [$I$ > 2 $\sigma(I)$]$^a$ | 0.1436 |

Considering that all family members of $Ba_6RE_2Ti_4O_{17}$ (RE=Nd-Yb) share the same crystal structure, as a representative, the schematic crystal structure of $Ba_6Nd_2Ti_4O_{17}$ is presented in Figure 2a. The framework of crystal structure is constructed by the corner-sharing connections of distorted $NdO_6$ octahedron with the $TiO_4$ tetrahedra and face-sharing $Ti_2O_9$ bioctahedra. Within the *ab* plane, magnetic $Nd^{3+}$ ions are occupied on a perfect triangular lattice with an equilateral nearest-neigboring $Nd^{3+}$-$Nd^{3+}$ intralayer distance of $a$ = 5.993(4) Å, then the triangular-lattice planes are stacked with an "AA"-type sequence along the crystallographic $c$ axis with two different interlayer distances as shown in Figure 2b. Further checking the connections for the $Nd^{3+}$ ions within the triangular-lattice plane, the adjacent $NdO_6$ octahedra are linked to each other by corner-sharing $TiO_4$ tetrahedra and $TiO_6$ octahedra and there exist two different Nd-O-Ti-O-Nd superexchange routes, as shown in Figure 2c. It is noteworthy that the intralayer $Nd^{3+}$-$Nd^{3+}$ distance $d_0$ = $a$ = 5.993(4) Å of $Ba_6Nd_2Ti_4O_{17}$ is just between ~6.224 Å of triangular-lattice antiferromagnet $NdTa_7O_{19}$[18] and ~5.593 Å for triangular-lattice $NdMgAl_{11}O_{19}$,[11] then the intraplane exchange interactions of $Ba_6Nd_2Ti_4O_{17}$ should have comparable exchange interactions with the above two systems. Along the $c$-axis, the $Nd^{3+}$ planes are well separated by the nonmagnetic layers constructed by Ti-O/Ba-O polyhedra, but the superexchange pathways are different for the alternative $Nd^{3+}$ triangular layers. For the neighboring two $Nd^{3+}$ layers with distance $d_1$ = 7.426(3) Å, the interplane superexchange pathway are realized via the $Ti_2O_9$ dimers with the Nd-O-Ti-O-Ti-O-Nd bridges. While for the neighboring $Nd^{3+}$ layers with distance $d_2$ = 7.549(7) Å, they are separated by double sheets of $TiO_4$ tetrahedra, as illustrated in Figure 1c and Figure S2b. Accordingly, the Nd-Nd connections are realized through the "…Nd-$Ti_2O_9$-Nd-$TiO_4$/$TiO_4$-Nd-$Ti_2O_9$-Nd…"pathways along the $c$-axis. Further considering that the magnetic exchange interactions depend on their superexchange pathway, the interlayer exchange interactions between the neighboring Nd-Nd layers with space separation of $d_2$ are much smaller than to the ones with distance of $d_1$, due to the absence of co-sharing oxygen atoms between the double sheets of $TiO_4$ tetrahedra (see Figure S2b in supporting information), which breaks the equivalency of interlayer exchange interactions for the alternative neighboring $Nd^{3+}$ triangular layers. Thus, $Ba_6Nd_2Ti_4O_{17}$ compound can be regarded as the layered triangular lattice system with two alternative interlayer magnetic exchange interactions.

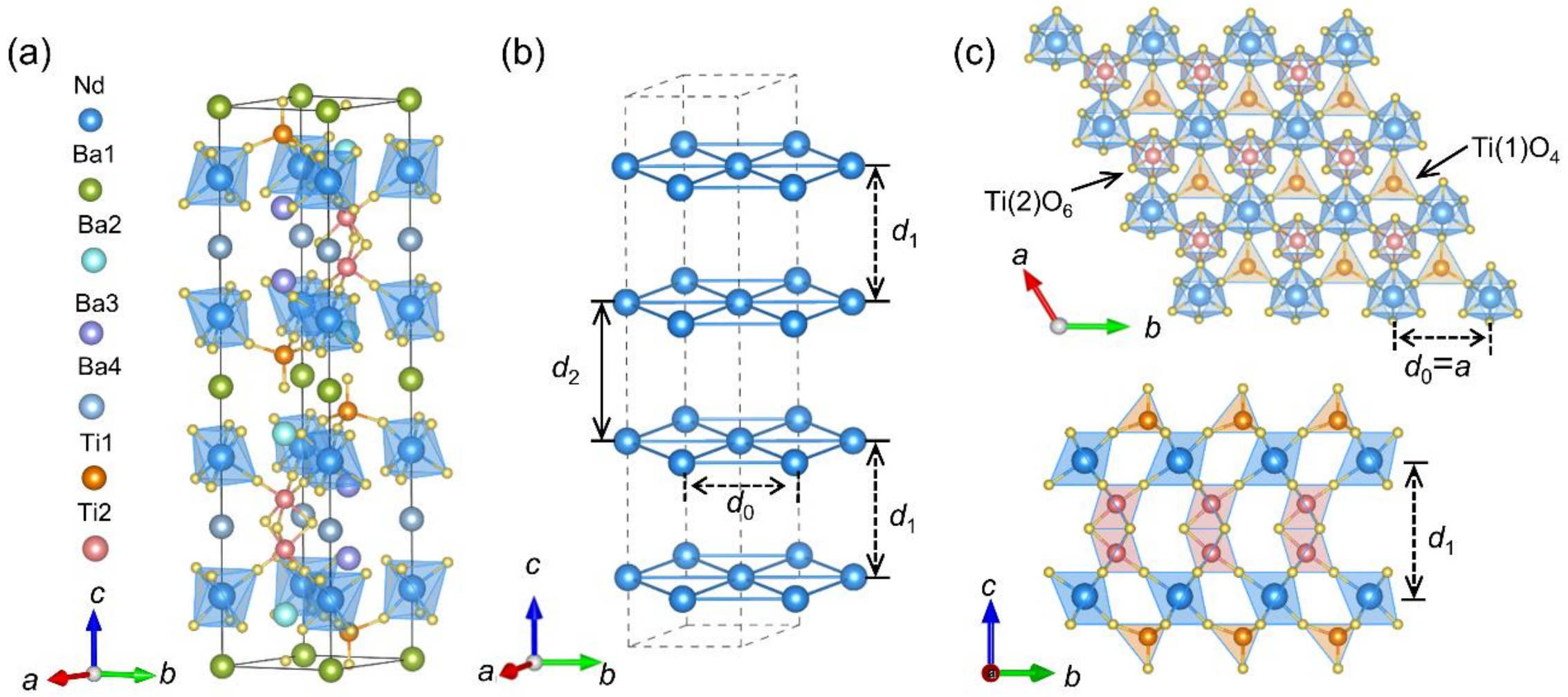

**Figure 2**. (a) The schematic crystal structure of $Ba_6Nd_2Ti_4O_{17}$ in the unit cell. (b) The lattice geometry of magnetic $Nd^{3+}$ triangular-lattice planes and its stacking sequence along the *c*-axis, the in-plane distance of $Nd^{3+}$ ions is denoted by $d_0$ and the interplanar distances are denoted by $d_1$ and $d_2$. (c) Top view (upper panel) and side view (lower panel) of the $NdO_6$ octahedra with the connections of $TiO_4$ and $TiO_6$ groups.

The unit cell of $Ba_6Nd_2Ti_4O_{17}$ consists of eleven crystallographic sites: one Nd atom (Wyckoff site 4*e*), four Ba atoms (Wyckoff site 2*a*, 2*b*, and 4*f*), two Ti atoms (Wyckoff site 4*f*), and four O atoms (Wyckoff site 4*f*, 6*h*, and 12*k*), the detailed coordination environments of Nd, Ti1, Ti2, Ba1, Ba2, Ba3, Ba4 with surrounding oxygen atoms are depicted in Figure S1a. The local coordinate environment of $NdO_6$ octahedra has point group symmetry $D_{3d}$ analogous to the case of $KCeO_2$ and $NaYbO_2$.[23,24] Based on the structure analysis of $Ba_6Nd_2Ti_4O_{17}$ single crystal, no antisite occupancy between the magnetic $Nd^{3+}$ ions and non-magnetic $Ba^{2+}/Ti^{4+}$ cations in $Ba_6Nd_2Ti_4O_{17}$ are detected, which can be due to the significant difference of ionic radii of 0.983 Å, 1.35 Å and 0.605 Å for $Nd^{3+}$, $Ba^{2+}$ and $Ti^{4+}$ ions as well as their different coordination environments. For other $Ba_6RE_2Ti_4O_{17}$ samples, the structure refinements reveal only a tiny amount (0.06%–1.57%) of RE/Ti site disorder in our $Ba_6RE_2Ti_4O_{17}$ (RE = Sm, Gd, Dy-Yb) polycrystalline samples, suggests the weak antisite disorder for $Ba_6RE_2Ti_4O_{17}$ compounds. This weak disorder is attractive to investigate the intrinsic magnetism from the perfect triangular lattice of $RE^{3+}$ moments.

**Magnetic Properties of $Ba_6RE_2Ti_4O_{17}$.** Magnetic susceptibility $\chi(T)$ measurements for the $Ba_6RE_2Ti_4O_{17}$ samples were carried out in temperature range from 1.8 K to 300 K under field of $\mu_0H$ = 0.1 T, the results are presented in Figure 3. The inverse susceptibility $1/\chi(T)$ is fitted by the Curie−Weiss (CW) law $1/\chi = (T - \theta_{cw})/C$, where $C$ is the Curie Constant, $\theta_{cw}$ is the CW temperature. The effective moment $\mu_{eff}$ is calculated by $\mu_{eff} = (3k_BC/N_A)^{1/2}$, where $k_B$ is the Boltzmann constant and $N_A$ is Avogadro's constant. Considering that the thermal population of electrons occupied on different crystal electric field (CEF) levels are dependent on the temperature, the CW fits are performed at both high-temperature ($T$ = 120–300 K) and low-temperature ($T$ = 5–10 K) regions, respectively. The $\chi^{-1}(T)$ data and CW fitting at low temperature regions are presented in Figure S3. The obtained $\theta_{cw}$ and $\mu_{eff}$ from the CW fits are summarized in Table 3. As shown in Figure 4, the isothermal field-dependent magnetizations $M(\mu_0H)$ at different temperatures were fitted with the Brillouin function $B_J(x) = \frac{2J_{\rm eff}+1}{2J_{\rm eff}} coth \frac{2J_{\rm eff}+1}{2J_{\rm eff}} x - \frac{1}{2J_{\rm eff}} coth \frac{x}{2J_{\rm eff}}$, where $x = g_J \mu_B J_{\rm eff} \mu_0 H/k_B T$, $J_{\rm eff}$ is the effective angular momentum, $g_J$ is the Lande's factor, and $\mu_B$ is the Bohr magneton. To perform the Brillouin function fits, the values of $J_{\rm eff}$ are fixed on the $M(\mu_0 H)$ curves at different temperatures for $Ba_6RE_2Ti_4O_{17}$ (RE=Nd, Gd, Dy, Ho, Er, Yb), the *g*-factor is used as adjusting parameter to fit the experimental data. Additionally, the Lande *g*-factors for some of compounds are determined based on the X-band electron spin resonance (ESR) results using $g_J = h\nu/\mu_B H_r$ (*h* is Plank's constant; $\nu$ = 9.4 GHz is the microwave frequency; $H_r$ is the resonance field). The elementary magnetic properties for $Ba_6RE_2Ti_4O_{17}$ are described separately as below.

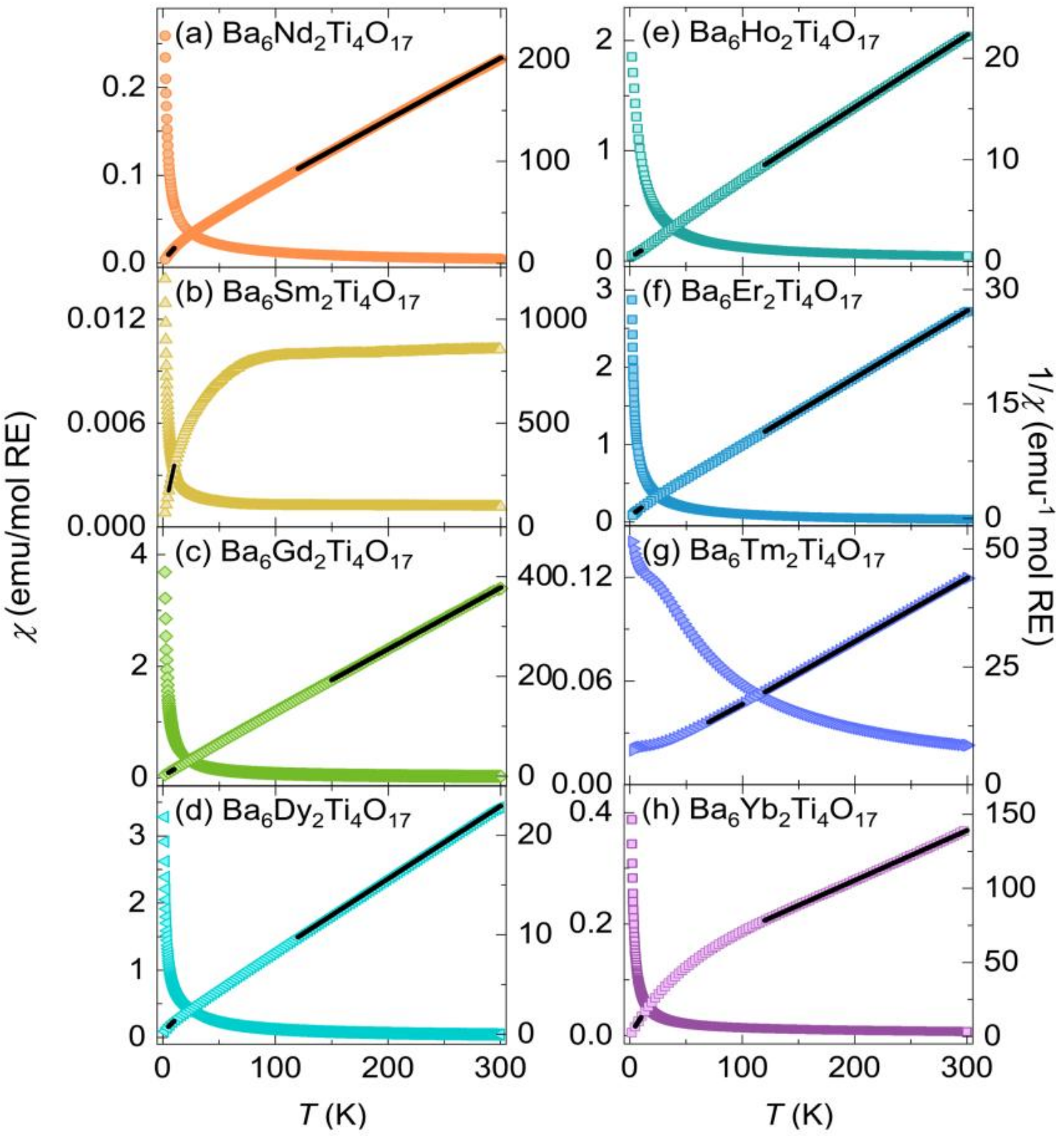

**Figure 3**. Temperature dependence of magnetic susceptibility $\chi(T)$ and inverse susceptibility $\chi^{-1}(T)$ measured at $\mu_0 H$ = 0.1 T for $Ba_6RE_2Ti_4O_{17}$ (RE = Nd, Sm, Gd, Dy-Yb), respectively. The black solid lines show the Curie-Weiss fitting in high temperature and low temperature regimes.

**$Ba_6Nd_2Ti_4O_{17}$.** Figure 3a shows the magnetic susceptibility $\chi(T)$ curve of $Ba_6Nd_2Ti_4O_{17}$, no magnetic order is detected with temperature down to 1.8 K. The CW fits on $\chi^{-1}(T)$ at high temperatures yield CW temperature $\theta_{cw}$ = −33.7 K and effective moment $\mu_{\rm eff}$ = 3.64 $\mu_B/Nd^{3+}$. This magnetic moment is consistent with the expected value $g_J[J(J+1)]^{1/2}$ = 3.62 $\mu_B/Nd^{3+}$ for free $Nd^{3+}$ ($^4f^3$,$^4I_{9/2}$) ions. As decreased temperatures, magnetic contribution from the electrons at excited CEF levels is reduced which can be responsible for the variation of the slope of $\chi^{-1}(T)$ curve. The low-temperature fits give $\theta_{cw}$ = −1.74 K and $\mu_{\rm eff}$ = 2.50 $\mu_B/Nd^{3+}$. The smaller $\mu_{\rm eff}$ is expected due to the CEF effect, where more population of electrons will occupy on the lowest-lying Kramers doublet states at the low temperatures.[25] The magnetization $M(\mu_0 H)$ curves of $Ba_6Nd_2Ti_4O_{17}$ at different temperatures are depicted in Figure 4a. The experimental saturated magnetization $M_S$~1.25 $\mu_B$ /Nd at 2 K is close to half value of the effective magnetic moment, similar to the report in other $Nd^{3+}$-based oxides.[25-27] The fitting of the Brillouin function for magnetization curves at 2 K gives a good fit with the experimental data, which is due to the weak magnetic exchange interactions between $Nd^{3+}$ moments.

**$Ba_6Sm_2Ti_4O_{17}$**. The susceptibility $\chi(T)$ of $Ba_6Sm_2Ti_4O_{17}$ is shown in Figure 3b. At high temperatures, it exhibits a temperature independent Van Vleck contribution from $Sm^{3+}$ ($4f^5$,${}^6H_{5/2}$) ions as report in many Sm-containing compounds.[26,28] Thus, the CW fitting is y performed at low temperatures between 5 and 10 K, which results in $\mu_{eff}$ = 0.567 $\mu_B$/Sm and $\theta_{cw}$ = −1.98 K, respectively. The negative $\theta_{cw}$ indicates the antiferromagnetic interactions between $Sm^{3+}$ moments. The $\mu_{eff}$ value is smaller than the value 0.83 $\mu_B$/Sm of free ion but similar to the ones reported in $Sm^{3+}$-containing oxides, such as $Sm_3Sb_3Mg_2O_{14}$ (0.53 $\mu_B$/Sm) and $Sm_2Zr_2O_7$ (0.50 $\mu_B$/Sm).[26,28] The reduced moments in these compounds may be related to the CEF splitting of $J$ = 5/2 multiplet of $Sm^{3+}$ ions. The field-dependent magnetization $M(\mu_0H)$ at 2 K shows a nearly field dependence without magnetic saturation up to $\mu_0H$ = 7 T (see Figure 4b).

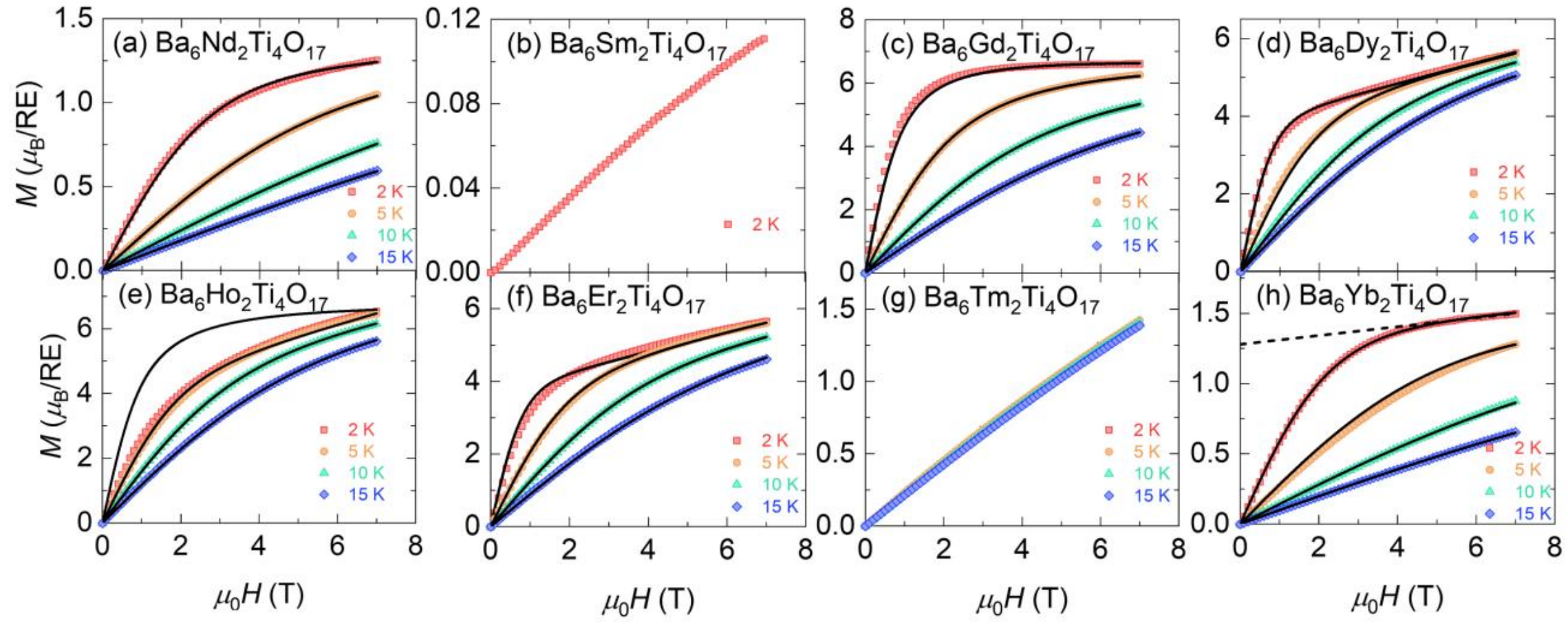

**Figure 4**. The isothermal field-dependent magnetization $M(\mu_0H)$ curves at different temperatures for $Ba_6RE_2Ti_4O_{17}$ (RE = Nd, Sm, Gd, Dy-Yb), the black solid lines show the fitted curves by Brillouin function.

**$Ba_6Gd_2Ti_4O_{17}$**. The $\chi(T)$ of $Ba_6Gd_2Ti_4O_{17}$ is shown in Figure 3c, which doesn't show any long-range magnetic order down to 1.8 K. High temperature CW fitting of $1/\chi(T)$ yields $\theta_{cw}$ = − 6.4 K and $\mu_{eff}$ = 8.05 $\mu_B$, the fitted $\mu_{eff}$ is consistent with the value of $g_J[J(J+1)]^{1/2}$ = 7.94 $\mu_B/Gd^{3+}$ for the free $Gd^{3+}$ ($4f^7$,$S$=7/2, $L$=0) ions . The low-temperature fitting gives $\theta_{cw}$ = -0.276 K and $\mu_{eff}$ = 7.69 $\mu_B/Gd^{3+}$, here the negative $\theta_{cw}$ value indicates the antiferromagnetic interactions between $Gd^{3+}$ ions. The $M(\mu_0H)$ curves of $Ba_6Gd_2Ti_4O_{17}$ at 2 K display a nonlinear field dependence behavior and reaches saturation at field~4 T with $M_S$ ~6.61 $\mu_B$ (Figure 4c), which is close to the saturated magnetization $M_{sat} = g_J J \mu_B$ = 7$\mu_B/Gd^{3+}$ of $Gd^{3+}$ moments with Heisenberg anisotropy of $Gd^{3+}$ spins. The Brillouin function gives the well-matched fits to the experimental $M(\mu_0H)$ data using $J_{eff}$ = 7/2, indicative of the dominant paramagnetic behaviors at 2 K. From the ESR spectra shown in Figure 5a, single resonance lines can be identified in the measured temperature regions. As temperature decreases from 50 K to 2 K, the $g$-factor shows a gradual increase from ~2.07 to ~2.49 as shown in Figure 6a, possibly due to the development of short-range AFM correlations.

**$Ba_6Dy_2Ti_4O_{17}$**. Figure 3d shows the $\chi(T)$ curve of $Ba_6Dy_2Ti_4O_{17}$. The high temperature CW fits yield $\theta_{cw}$ = −13.3 K and $\mu_{eff}$ = 10.4 $\mu_B$/Dy, this moment is close to 10.6 $\mu_B$/Dy of free $Dy^{3+}$ ($^6H_{15/2}$) ions. At low temperatures, linear fit on $\chi^{1}(T)$ gives $\mu_{eff}$ = 8.49 $\mu_B$/Dy and $\theta_{cw}$ = −1.85 K. The negative $\theta_{cw}$ indicates AFM couplings between local $Dy^{3+}$ spins. Figure 4d shows the $M(\mu_0H)$ curves at different temperatures. At 2 K, the magnetization shows nonlinear field dependence at field below 1.5 T and has linear field dependence at higher fields, which can be possibly from the Van Vleck magnetic contribution. By including a linear-field contribution into the Brillouin function, we can well fit the experimental $M(\mu_0H)$ curves. The maximum magnetization value $M_{sat}$ is 5.63 $\mu_B$/Dy in an applied field of $\mu_0H$ = 7 T, which is close to half of the expected $Dy^{3+}$ saturation magnetization of $g_JJ\mu_B$ = 10 $\mu_B/Dy^{3+}$ with Ising-like magnetic anisotropy. From the ESR spectra measured at different temperatures (see Figure 5b), the calculated $g$-factors increase and reach the values of 6.3- 6.7 as temperature is below 5 K, as presented in Figure 6b.

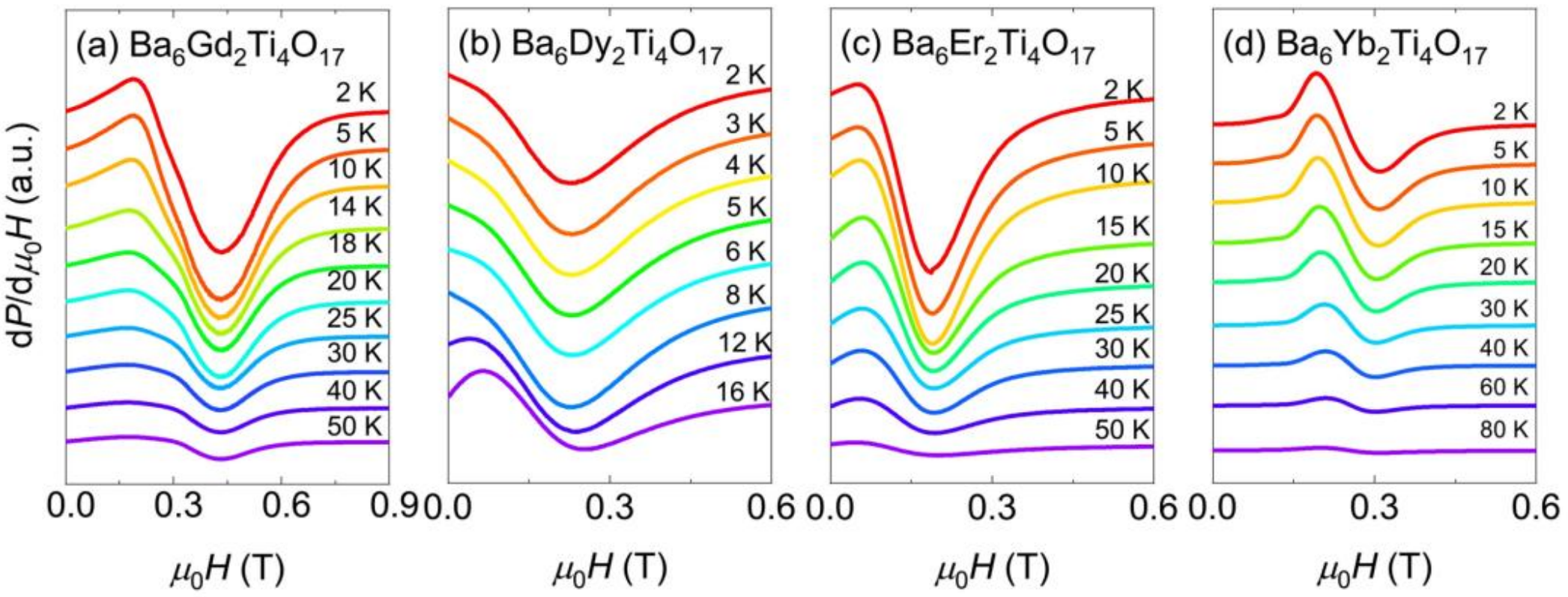

**Figure 5**. The ESR spectra at selected temperatures for $Ba_6RE_2Ti_4O_{17}$ (RE = Gd, Dy, Er, Yb) compounds with (a) RE = Gd, (b) RE = Dy, (c) RE = Er, (d) RE = Yb.

**$Ba_6Ho_2Ti_4O_{17}$**. As displayed in Figure 3e, the $\chi(T)$ curve of $Ba_6Ho_2Ti_4O_{17}$ does not show any hints of long-range magnetic order down to 1.8 K. The temperature-dependent susceptibility is similar to the one of $Ba_6Dy_2Ti_4O_{17}$. The high temperature CW fits give $\mu_{eff}$ =10.6 $\mu_B$/Ho and $\theta_{cw}$=−14.3 K, and low-temperature analysis yield $\theta_{cw}$ = −4.83 K and $\mu_{eff}$ = 10.5 $\mu_B/Ho^{3+}$. The obtained moment value is in accordance with the moment 10.60 $\mu_B/Ho^{3+}$ of free $Ho^{3+}$ ($^5I_8$) ions. Considering that the $Ho^{3+}$ ion in $Ba_6Ho_2Ti_4O_{17}$ share the same $D_{3d}$ symmetry with the pyrochlore-lattice $Ho_2Ti_2O_7$ and triangular-lattice $KBaHo(BO_3)_2$ magnets,[29,30] the formation of similar non-Kramers doublet with easy-axis (Ising) anisotropy is expected. The determination of CEF splitting using inelastic neutron spectroscopy (INS) will be required to clarify this issue. The $M(\mu_0H)$ curve at 2 K exhibits maximum magnetization $M_S$ = 6.53 $\mu_B$/Ho at $\mu_0H$ = 7 T (Figure 4e), which is much smaller than $M_S$ = $g_JJ\mu_B$ = 10 $\mu_B/Ho^{3+}$ for free $Ho^{3+}$ ions. At $T \geq 5$ K, the experimental $M(\mu_0H)$ data can be well fitted by Brillouin function as expected for the paramagnetic state. In contrast, the Brillouin function fits deviate from the experimental data at 2 K, possibly due to the development of AFM interactions or the formation of non-Kramers doublet states at low temperatures.

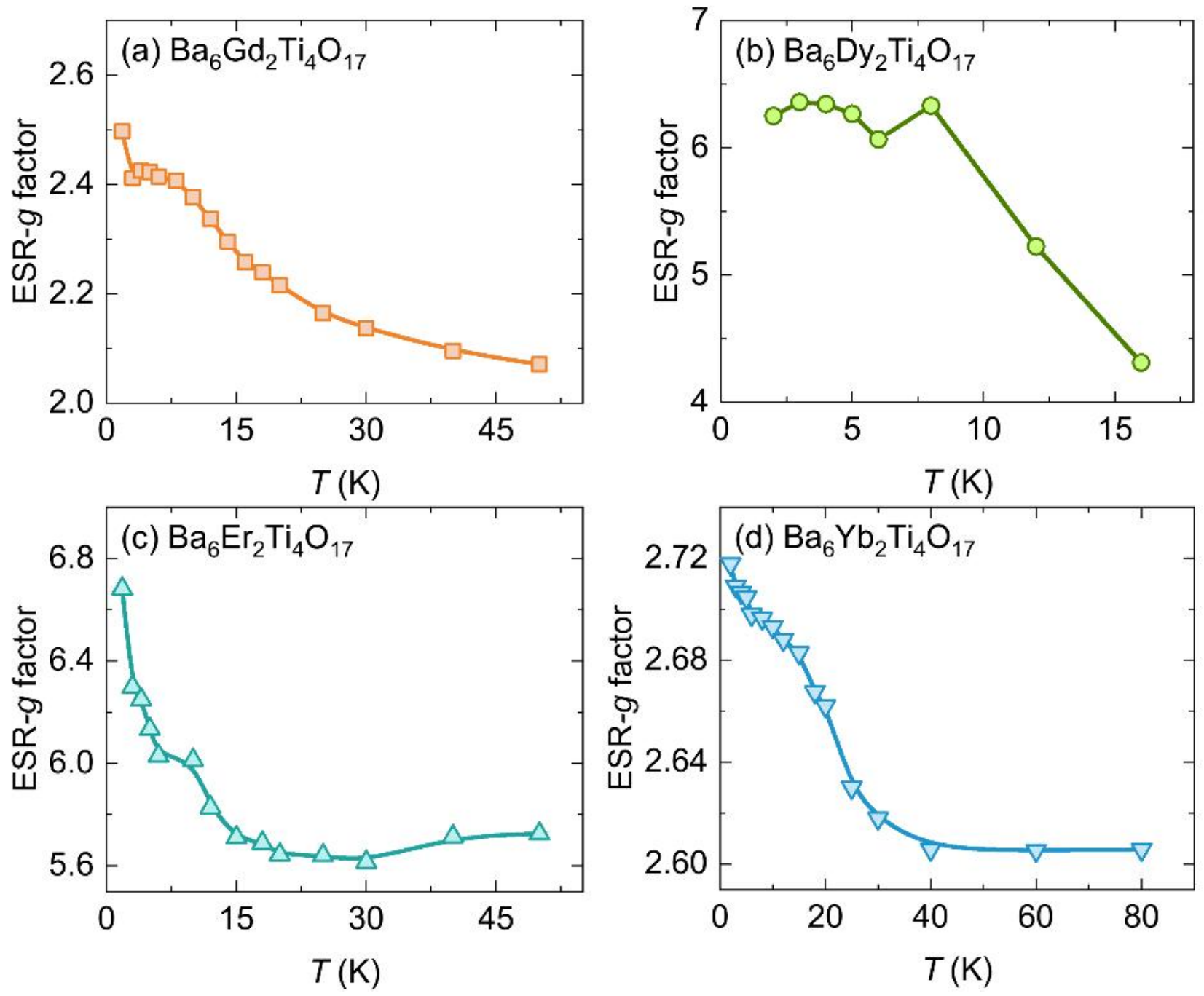

**Figure 6**. Temperature dependence of estimated ESR *g*-factors of $Ba_6RE_2Ti_4O_{17}$ compounds with (a) RE = Gd, (b) RE = Dy, (c) RE = Er, (d) RE = Yb.

**$Ba_6Er_2Ti_4O_{17}$**. The $\chi(T)$ curves of $Ba_6Er_2Ti_4O_{17}$ show no sign of magnetic transition down to 1.8 K, as shown in Figure 3f. At high-temperature regimes (120-300 K), the CW fits yield $\theta_{cw}$ = −9.46 K and $\mu_{eff}$ = 9.53 $\mu_B$/Er close to the 9.59 $\mu_B$ of free $Er^{3+}$ ($4f^{11}$, $J$=15/2) ions. The low temperature fits give $\theta_{cw}$ = −2.08 K and $\mu_{eff}$ = 8.30 $\mu_B$/Er. This obtained $\theta_{cw}$ is smaller than the observation in $ErMgGaO_4$ but close to the $\theta_{cw}$ values in Er chalcogenides $AErSe_2$ (A = $Na^+$, $K^+$) and $K_3Er(VO_4)_2$.[12,31,32] Figure 4f presents the isothermal $M(\mu_0H)$ curves at different temperatures for $Ba_6Er_2Ti_4O_{17}$, which exhibit a nonlinear relation at low fields and then increase linearly above 4 T with maximum magnetization $M_S$~5.65 $\mu_B$/Er at 7 T. From the ESR spectra shown in Figure 5c, single resonance lines are identified at different measured temperatures. As shown in Figure 6c the *g*-factor shows a steep increase from 5.72 to 6.68 as temperature decreases from 50 K to 2 K. If using the obtained g-factor $g_1$=6.68 and $J_{eff}$=1/2 at 2 K, the estimated magnetization $M_{sat} = gJ_{eff}\mu_B \sim 3.34\mu_B$ is much smaller the experimental value, which suggests the existence of magnetic contribution from the orbital moments of $Er^{3+}$ ions.

**$Ba_6Tm_2Ti_4O_{17}$**. The magnetic susceptibility of $Ba_6Tm_2Ti_4O_{17}$ is shown in Figure 3g. As decreased temperatures, the $\chi(T)$ displays a broad hump at ~15 K, possibly related to the formation of two low-lying nonmagnetic singlet states for $Tm^{3+}$ ($4f^{12}$) ions in the local $D_{3d}$ symmetry environments analogous to the case of $Ba_9Tm_2(SiO_4)_6$ and $TmMgGaO_4$ compounds.[21,33] High temperature CW fits give $\theta_{cw}$ = −25.7 K and $\mu_{eff}$ = 7.71 $\mu_B$/Tm, the moment value is close to the $\mu_{eff}$ =7.57 $\mu_B$ for $Tm^{3+}$ free ions. At low temperatures, the isothermal magnetization $M(\mu_0H)$ curves don't show any sign of saturation to 7 T as shown in Figure 4g, this is also in accordance with the formation of two low-lying singlets state, which can give a Van Vleck magnetic contribution. At $T$ = 2 K, the maximum magnetization value ~1.4 $\mu_B$ at $\mu_0H$ = 7 T is far smaller than the expected saturated magnetization $M_S = g_J J \mu_B = 7\mu_B$/$Tm^{3+}$ for free $Tm^{3+}$ ions.

**$Ba_6Yb_2Ti_4O_{17}$.** As shown in Figure 3h, high temperature CW fits on the $\chi^{-1}(T)$ of $Ba_6Yb_2Ti_4O_{17}$ give $\theta_{cw}$ = −112.6 K and $\mu_{eff}$ = 4.86 $\mu_B$/Yb, the moment is close to the value of free $Yb^{3+}$ ion (4.53 $\mu_B$). The low temperature CW fits give $\theta_{cw}$ = −0.71 K and $\mu_{eff}$ = 2.57 $\mu_B$/Yb. The low-temperature fitted negative $\theta_{cw}$ indicates a dominant antiferromagnetic exchange interaction between $Yb^{3+}$ moments. The $M(\mu_0 H)$ curves at selected temperatures are shown in Figure 4h. After subtracting a linear dependent magnetization contribution (the slope of the linear fit for $H$ > 4 T is $\chi_{vv}$ = 1.75×10$^{-2}$ emu/mol), the saturated moment is ~1.28 $\mu_B$/$Yb^{3+}$. This small magnetization supports the formation of doublet ground states described by $J_{eff}$ = 1/2 effective moments. Also, the Brillouin function fitting yields a powder-average Landé $g$ factor 2.67 at 2 K, this is in consistent with $g$ = 2.7 evaluated from the single ESR spectra as shown in Figure 5d. The ESR $g$-factor increases dramatically as temperature decreases below 30 K (see Figure 6d). Using the obtained ESR $g$-factors at 2 K, the estimated effective magnetic moment $\mu_{eff}$ = $g\sqrt{J(J+1)}\mu_B$ = 2.34 $\mu_B$ is close to the experimental data.

For the compounds containing $RE^{3+}$ ions, they can be divided into two categories: Kramers ions with odd 4$f$ electrons ($Nd^{3+}$, $Sm^{3+}$, $Dy^{3+}$, $Er^{3+}$, and $Yb^{3+}$) and non-Kramers ions with even numbers of 4$f$ electrons ($Ho^{3+}$, and $Tm^{3+}$). Among them, $Gd^{3+}$ is special and has half-filled 4f shell ($4f^7$, $S$=7/2, $L$= 0) with Heisenberg-like anisotropy and effective moment $S_{eff}$=7/2, then the negative value of $\theta_{cw}$= −0.296 K indicates the AFM interactions for $Ba_6Gd_2Ti_4O_{17}$. For the other compounds containing Kramers ions, low temperature magnetism can be well described by effective moment $J_{eff}$ =1/2 protected by time-reversal symmetry. While, for non-Kramers $RE^{3+}$ ions, the situation is complicated and it can be taken as effective pseudospin $S_{eff}$ = 1/2 state at low temperatures in case of the formation of a pair of low-lying CEF ground state singlet, which depends on the point group symmetry of $REO_n$ coordinate environments. Here, the energy of magnetic interactions can be evaluated by the mean-field approximations of $Ba_6RE_2Ti_4O_{17}$, the superexchange interaction ($J_{nn}$) between localized $RE^{3+}$ moments can be approximated by the relation,[34] $J_{nn} = (3k_B\theta_{cw})/zS(S+1)$, where $S$ represents the effective spin quantum number and $z$ is the number of nearest-neighbor spins (here $z$ = 6). The dipolar interaction energy ($D$) can be estimated by $D = \mu_0\mu_{eff}^2 / [4\pi(R_{nn})^3]$, where $R_{nn}$ is the space separation between nearest-neighbor $RE^{3+}$ ions within the triangular-lattice plane. Since the mean field approximations don't consider the local single-ion anisotropy of $RE^{3+}$ ions and other contributions of exchange interaction, it only can give the rough estimation on the dipolar interaction and superexchange interaction in the energy scale, the accurate determination requires the inelastic neutron spectroscopy (INS) measurements on the $Ba_6RE_2Ti_4O_{17}$ compounds in the future. Moreover, the low-temperature fitted magnetic parameters $\theta_{cw}$ and $\mu_{eff}$ can relatively well reflect the magnetic interactions of ground state, then $J_{nn}$ and $D$ values are calculated using the low-$T$ fitted $\theta_{cw}$ and $\mu_{eff}$, as also provided in Table 3. From that, $Ba_6Yb_2Ti_4O_{17}$ as an example, the intralayers' exchange interaction and dipolar interaction is $D$~0.020 K and $J_{nn}$~-0.47 K, and here the distance ratio for the $Yb^{3+}$ ion between the interlayer and intralayer ones $\delta$ = $d_1/d_0$~1.25, which is larger than ~1.17 for $YbBO_3$ compound.[35] Moreover, the estimated $J/k_B$ (~0.47 K) in $Ba_6Yb_2Ti_4O_{17}$ is comparable to that of $YbBO_3$ (~0.53 K). As another typical feature, the "AA" type stacking fashion of layered triangular-lattice in $Ba_6Yb_2Ti_4O_{17}$ doesn't introduce the additional geometric frustration from the stacking order, this is different from the "ABC" stacking fashion in $NaBaYb(BO_3)_2$ and $NaYbS_2$ compounds.[36,37]

Table 3. The obtained magnetic parameters $\theta_{CW}$ and $\mu_{eff}$ from the Curie–Weiss fitting of $\chi(T)$ for $Ba_6RE_2Ti_4O_{17}$ (RE = Nd, Sm, Gd, Dy-Yb) samples and the effective moment $\mu_{fi}$ of free ions calculated by $g[J(J+1)]^{1/2}$, the calculated superexchange interaction ($J_{nn}$) and the in-plane dipolar interaction ($D$).

| RE | High *T* fit | $\theta_{cw}$ (K) | $\mu_{eff}$ ($\mu_B$) | Low *T* fit | $\theta_{cw}$ (K) | $\mu_{eff}$ ($\mu_B$) | $\mu_{fi}$ ($\mu_B$) | *D* (K) | $J_{nn}$ (K) using *J* | $J_{nn}$ (K) using $J_{eff}$ = 1/2 |
|---|---|---|---|---|---|---|---|---|---|---|
| **Nd** | 120-300 K | -33.7 | 3.64 | 5-10 K | -1.74 | 2.50 | 3.62 | 0.018 | -0.035 | -1.16 |
| **Sm** | – | – | – | 5-10 K | -1.98 | 0.567 | 0.85 | 0.0009 | -0.113 | -1.32 |
| **Gd** | 120-300 K | -6.40 | 8.05 | 5-10 K | -0.276 | 7.69 | 7.94 | 0.173 | -0.009 | N/A |
| **Dy** | 120-300 K | -13.3 | 10.4 | 5-10 K | -1.85 | 8.49 | 10.6 | 0.213 | -0.015 | -1.23 |
| **Ho** | 120-300 K | -14.3 | 10.6 | 5-10 K | -4.83 | 10.5 | 10.6 | 0.328 | -0.034 | -3.22 |
| **Er** | 120-300 K | -9.46 | 9.53 | 5-10 K | -2.08 | 8.30 | 9.59 | 0.205 | -0.016 | -1.39 |
| **Tm** | 120-300 K | -25.7 | 7.71 | 70-100 K | -34.2 | 7.92 | 7.57 | 0.188 | -0.407 | -22.8 |
| **Yb** | 120-300 K | -112.6 | 4.86 | 5-10 K | -0.71 | 2.57 | 4.53 | 0.020 | -0.022 | -0.47 |

## Magnetic Property and Specific Heat of $Ba_6Nd_2Ti_4O_{17}$ Single Crystal.

Temperature dependence of magnetic susceptibilities of $Ba_6Nd_2Ti_4O_{17}$ single crystals were measured under field of $\mu_0H$ =0.1 T. Figure 7a shows the measured $\chi(T)$ curves for field along *c*-axis ($\mu_0H$ //*c*) and perpendicular to *c*-axis ($\mu_0H$ // *ab*-plane), respectively. As shown in Figure 7b, high temperature fitted effective moments $\mu_{eff,ab}$= 3.46 $\mu_B$/Nd ($\mu_0H$ // *ab*) and $\mu_{eff,c}$= 3.82 $\mu_B$/Nd ($\mu_0H$ //*c*) are close to the value for free $Nd^{3+}$ ion (3.62 $\mu_B$). Using the magnetic susceptibilities of single crystals, the calculated average susceptibility ($\bar{\chi}$) of powders by $\bar{\chi} = \frac{1}{3}\chi_c + \frac{2}{3}\chi_{ab}$ agrees well with the experimental $\chi(T)$ data. For $Ba_6Nd_2Ti_4O_{17}$, the typical feature is that its magnetic susceptibilities exhibit the Ising-like anisotropy with easy magnetization along *c*-axis, the spin anisotropy ($\chi_c/\chi_{ab}$) reaches ~7.5 at 2 K. The low-temperature (2−4 K) fitted $\theta_{cw,c}$= −0.19 K and $\theta_{cw,ab}$= −1.2 K further support easy-axis anisotropy with dominant AFM exchange interactions, this is quite similar to the report in triangular-lattice $NdTa_7O_{19}$ with dominant Ising-like character,[18] but different from the triangular-lattice $TbInO_3$ with an in-plane XY-type anisotropy.[38] The isothermal magnetization $M(\mu_0H)$ curves were measured at selected temperatures, the experimental results for $\mu_0H$ //*c* and $\mu_0H$ // *ab* are shown in Figure 7c, d, respectively. Along the *c*-axis, the magnetization reaches saturation with $M_S$~ 1.79 $\mu_B$/Nd at 7 T, which is close to 3 times larger than the value ~0.64 $\mu_B$/Nd for $\mu_0H$ // *ab*-plane. To check the magnetic anisotropy, the ESR spectra were measured on $Ba_6Nd_2Ti_4O_{17}$ single crystals along the above two directions, the obtained ESR spectra at 2 K are shown in the inset of Figure 7e and 7f, respectively. We can identify two resolved ESR lines with peaks at $\mu_0H_{r1}$ and $\mu_0H_{r2}$ denoted by the dashed lines, which can be due to the hyperfine structure of $^{143}$Nd (I = 7/2) and $^{145}$Nd (I = 7/2) isotopes.[39,40] Based on the ESR spectra shown in Figure S4, the temperature-dependent average Lande *g*-factors ($g_{ave}$) are shown in Figure 7e and Figure 7f. The values of $g_{ab}$= 1.29 ($\mu_0H$ // *ab*) and $g_c$= 3.49 ($\mu_0H$ // *c*) at 2 K support a strong Ising-like magnetic anisotropy.

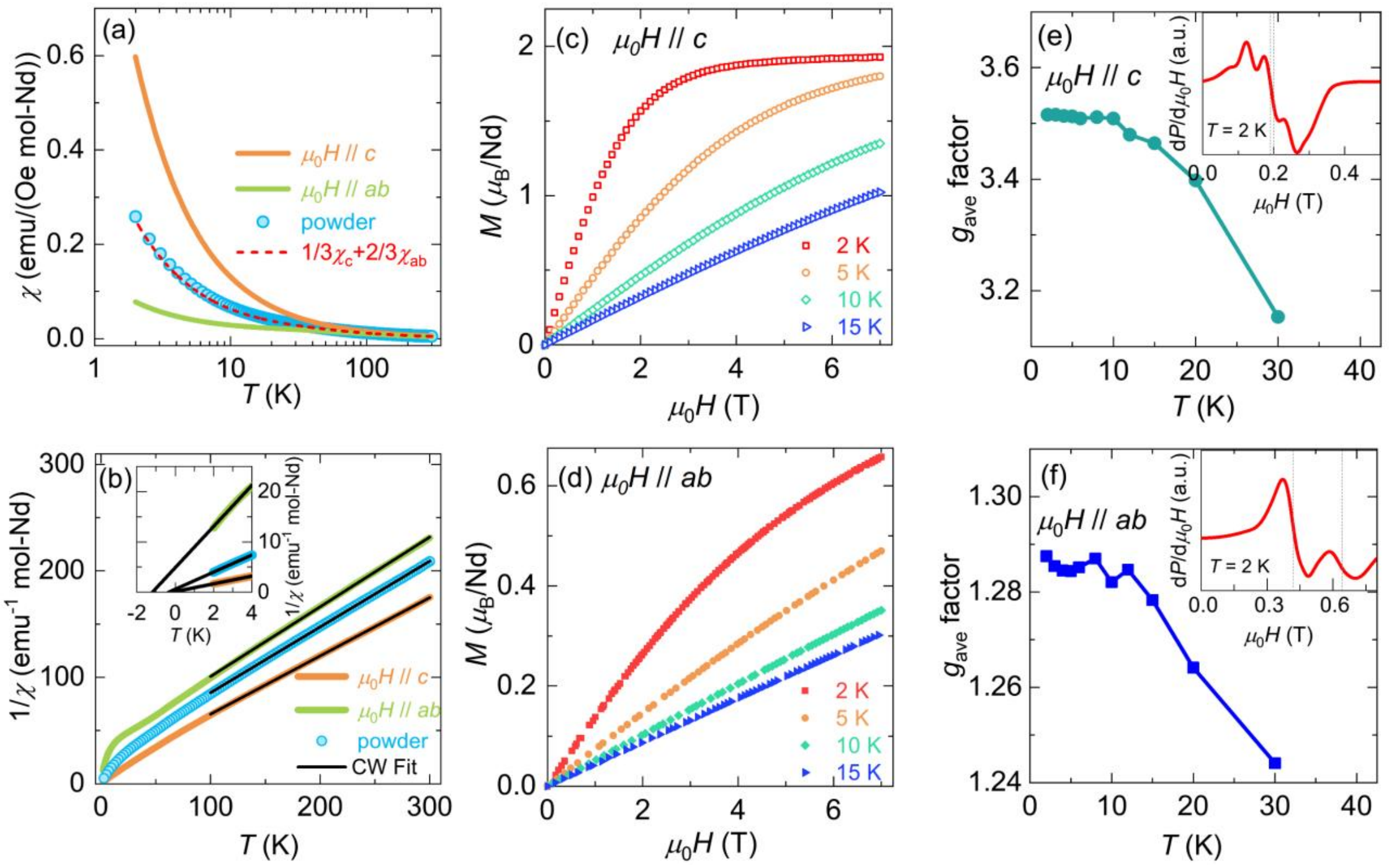

**Figure 7.** (a) Magnetic susceptibility ($\chi_c, \chi_{ab}$) of $Ba_6Nd_2Ti_4O_{17}$ single crystals for $\mu_0 H$ // *c*-axis and $\mu_0 H$ // *ab*-plane under field of 0.1 T, the susceptibility of polycrystals is also presented for comparison. The dashed red line shows the calculated magnetic susceptibility by $\bar{\chi} = \frac{1}{3}\chi_c + \frac{2}{3}\chi_{ab}$. (b) The Inverse susceptibility of $Ba_6Nd_2Ti_4O_{17}$ single crystals and its comparison with the ones of polycrystals, the black lines represent the CW fitted curves, the inset shows the low-temperature CW fitting analysis. (c, d) The isothermal $M(\mu_0 H)$ curves of $Ba_6Nd_2Ti_4O_{17}$ single crystal for $\mu_0 H$ // *c*-axis and $\mu_0 H$ // *ab*-plane, respectively. (e, f) The temperature dependence of *g*-factors obtained from the ESR spectra of $Ba_6Nd_2Ti_4O_{17}$ single crystal for $\mu_0 H$ // *ab*-plane and $\mu_0 H$ // *c*-axis, the inset show the ESR spectra at 2 K.

To unveil the magnetic ground state, the specific heat $C_p(T)$ measurements were performed on the $Ba_6Nd_2Ti_4O_{17}$ single crystals at different magnetic fields with $\mu_0 H$ // *c*-axis. Figure 8a shows the zero-field specific heat of $Ba_6Nd_2Ti_4O_{17}$. The zero-field $C_p(T)$ curves show a minimum at ~3 K then increase steeply down to the lowest measured temperatures. No any sharp peak is detected, ruling out the magnetic transition with temperature down to ~0.08 K. After subtracting the lattice contribution using the isostructural nonmagnetic $Ba_6Eu_2Ti_4O_{17}$ as the reference, temperature dependence of magnetic heat capacity $C_{mag}(T)$= $C_p(T)$-$C_{Latt}(T)$ is shown in Figure 8b. Under applied field, the $C_{mag}(T)$ curves don't show any sharp peak at $T \geq 0.1$ K indicating the absence of field-induced long-range magnetic order. For $\mu_0 H \geq 0.5$ T, the $C_{mag}(T)$ curves show a broad maximum and shift to higher temperatures. The height of the peaks with the same amplitude as $\mu_0 H \geq 1$ T reveals a Schottky-like anomaly. Further integrating the $C_{mag}(T)/T$ with respect to temperature, the changes of magnetic entropy ($\Delta S_{mag}$) at different fields are plotted on Figure 8c. As seen, the calculated magnetic entropy from 0.08 K to 15 K approaches $R\ln2$ for $\mu_0H$ ≥1 T, this is expected for the formation of $J_{eff}$ =1/2 Kramers doublet ground state. While the zero-field release of $\Delta S_{mag}$ only reaches ~0.27Rln2, leaving ~73% of entropy at temperatures below 0.08 K.

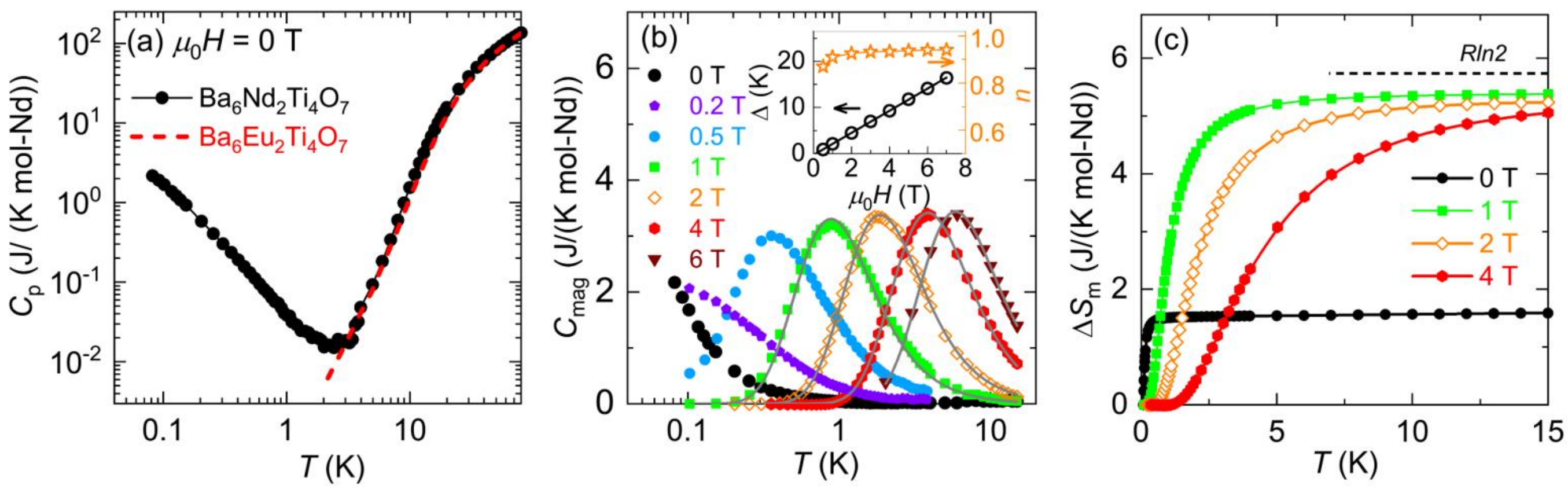

**Figure 8**. (a) Temperature dependence of zero-field specific heat $C_p(T)$ curves of $Ba_6Nd_2Ti_4O_{17}$ single crystal and $Ba_6Eu_2Ti_4O_{17}$ polycrystals. (b) Temperature dependence of magnetic specific heat $C_{mag}(T)$ of $Ba_6Nd_2Ti_4O_{17}$ single crystal at different fields for field along *c*-axis, the gray lines represent the fits by the two-level Schottky function, inset shows the energy gap $\Delta$ and the concentration *n* versus $\mu_0H$. (c) Magnetic entropy $\Delta S_m(T)$ of $Ba_6Nd_2Ti_4O_{17}$ under different fields.

Using the two-level Schottky function ($C_{sch}$), $C_{mag}$ ($T$) curves are fitted by the following relation: $C_{sch} = nR(\frac{\Delta}{T})^2 \frac{\exp(\Delta/T)}{[1+\exp(\Delta/T)]^2}$ , where *n* is the concentration of Schottky centers, and *R* is an ideal gas constant, and $\Delta$ is the Zeeman energy gap between two levels.[16] The fitted curves are denoted by the solid gray lines in Figure 8b. From the above fitting, the crystal field energy gap $\Delta$ increases linearly versus $\mu_0H$ (see the inset of Figure 8b), and the zero-field energy gap $\Delta$ is close to zero, the fitted $n \approx 1$ for $\mu_0H \geq 2$ T indicates the $Nd^{3+}$ spins are excited to the higher-energy levels in the two levels. As shown in Figure 8c, the released magnetic entropy $S_M(T)$ under $\mu_0H$ = 4 T reaches a saturation value ~$R\ln2$ in accordance with the doublet ground state. No further increase of $S_M(T)$ indicates a well separation for the first CEF excited level from the doublet ground state with a large energy gap. Thus, it can be well considered as an effective $J_{eff}$=1/2 system at low temperatures. The presence of magnetic frustration with frustration index $f = |\theta_{CW}/T_N|$ of 2.4~15 as well as its absence of long-range magnetic order down to 0.08 K. Compared to the lack of magnetic characterization of single crystal form on $NdTa_7O_{19}$ as Ising triangular-lattice QSL candidate,[18] here the successful growth of $Ba_6Nd_2Ti_4O_{17}$ single crystal allow us to establish its magnetic ground state and anisotropic magnetic behaviors.

## ■ CONCLUSIONS

A new family of rare-earth-based triangular lattice magnets, $Ba_6RE_2Ti_4O_{17}$ (RE = Nd, Sm, Gd, Dy−Ho), were successfully synthesized and magnetically characterized, where the layered magnetic triangular lattice planes are stacked in an eclipsed “AA”-type stacking fashion along the *c*-axis. Structural analysis reveals that these compounds are free of chemical antisite occupancy between the RE and Ba/Ti cations, making them attractive for studying the intrinsic triangular lattice physics. All $Ba_6RE_2Ti_4O_{17}$ compounds display dominant antiferromagnetic interactions between the $RE^{3+}$ moments without magnetic transition down to 1.8 K. The analyses of isothermal magnetizations reveal the presence of different magnetic anisotropy for different RE ions. More importantly, the synthesized $Ba_6Nd_2Ti_4O_{17}$ single crystals exhibit a strong Ising-like anisotropy along *c*-axis and absence of long-range magnetic order down to 0.08 K, being an Ising-type quantum magnet with effective $J_{eff}$=1/2 local moment.

## ■ Accession Codes

CCDC 2306558 contains the supplementary crystallographic data for this paper. This data can be obtained free of charge via http://www.ccdc.cam.ac.uk/data_request/cif, or emailing data_request@ccdc.cam.ac.uk.

## ■ AUTHOR INFORMATION

**Corresponding Author**

**Hanjie Guo –**Songshan Lake Materials Laboratory, Dongguan, Guangdong 523808, P. R. China; orcid.org/0000- 0001-6203-5557; Email: hjguo@sslab.org.cn

**Zhaoming Tian** − School of Physics and Wuhan National High Magnetic Field Center, Huazhong University of Science and Technology, Wuhan 430074, PR China. orcid.org/0000-0001-6538-3311; Email: tianzhaoming@hust.edu.cn.

**Authors**

**Fangyuan Song**–School of Physics and Wuhan National High Magnetic Field Center, Huazhong University of Science and Technology, Wuhan, 430074, P. R. China

**Andi Liu**–School of Physics and Wuhan National High Magnetic Field Center, Huazhong University of Science and Technology, Wuhan, 430074, P. R. China; Songshan Lake Materials Laboratory, Dongguan, Guangdong 523808, P. R. China

**Qiao Chen**–School of Physics and MOE Key Laboratory of Fundamental Physical quantum Physics, PGMF, Huazhong University of Science and Technology, Wuhan 430074, China

**Jin Zhou**–School of Physics and Wuhan National High Magnetic Field Center, Huazhong University of Science and Technology, Wuhan, 430074, P. R. China

**Jingxin Li**–Anhui Province Key Laboratory of Condensed Matter Physics at Extreme Conditions, High Magnetic Field Laboratory, HFIPS, Chinese Academy of Sciences, Hefei, Anhui 230031, P. R. China

**Wei Tong** –Anhui Province Key Laboratory of Condensed Matter Physics at Extreme Conditions, High Magnetic Field Laboratory, HFIPS, Chinese Academy of Sciences, Hefei, Anhui 230031, P.R. China

**Shun Wang**–School of Physics and MOE Key Laboratory of Fundamental Physical quantum Physics, PGMF, Huazhong University of Science and Technology, Wuhan 430074, China

**Yanhong, Wang**–Key Laboratory of Material Chemistry for Energy Conversion and Storage, School of Chemistry and Chemical Engineering, Huazhong University of Science and Technology, Wuhan, 430074, China.

**Hongcheng Lu**—Key Laboratory of Material Chemistry for Energy Conversion and Storage, School of Chemistry and Chemical Engineering, Huazhong University of Science and Technology, Wuhan, 430074, China.

**Songliu Yuan**−School of Physics, Huazhong University of Science and Technology, Wuhan 430074, PR China

## ■ ASSOCIATED CONTENT

### Supporting Information

The crystal structure parameters of $Ba_6RE_2Ti_4O_{17}$ polycrystals, low-temperature magnetic parameters and X-band ESR spectra of $Ba_6Nd_2Ti_4O_{17}$ single crystal. (PDF)

## ■ ACKNOWLEDGEMENTS

This work was supported by the National Natural Science Foundation of China (Grant No. 11874158), the Fundamental Research Funds of Guangdong Province (Grant No. 2022A1515010658) and Guangdong Basic and Applied Basic Research Foundation (Grant No.2022B1515120020). This work was supported by the synergetic extreme condition user facility (SECUF), and a portion of magnetic measurement was performed on the Steady High Magnetic Field Facilities, High Magnetic Field Laboratory. We would like to thank Guifen Ren for her assistance on the specific heat measurement and thank the staff of the analysis center of Huazhong University of Science and Technology for their assistance in structural characterizations.

## ■ REFERENCES

## For Table of Contents Only

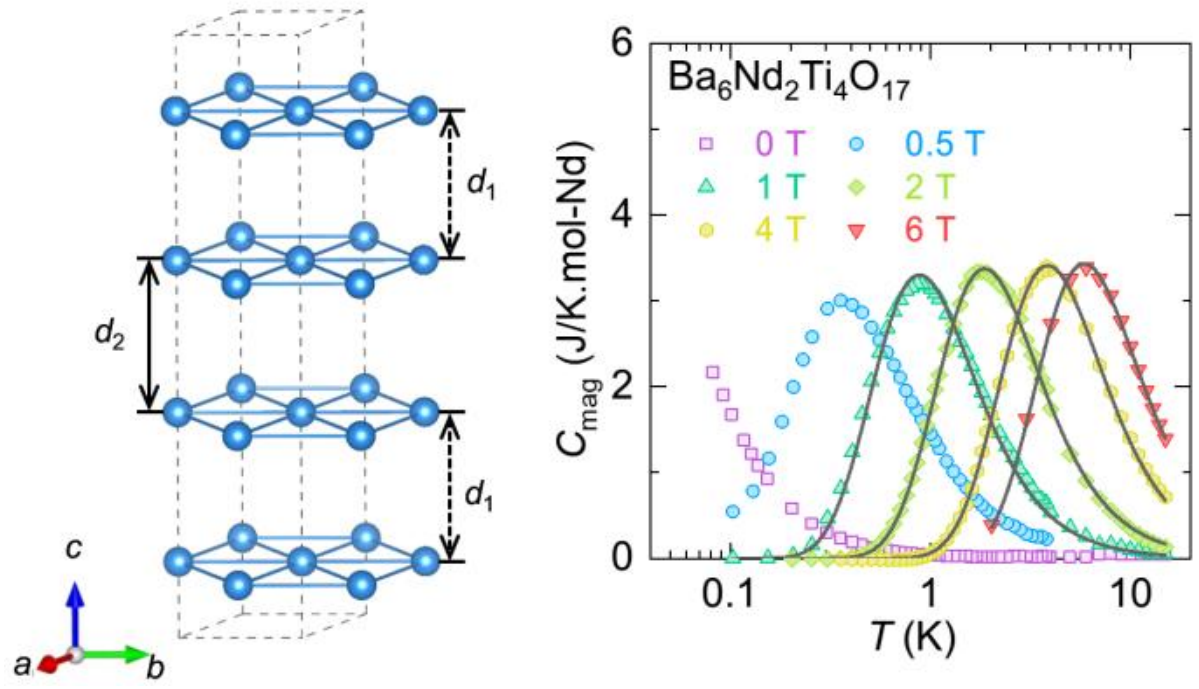

## Synopsis

The triangular-lattice magnet $Ba_6Nd_2Ti_4O_{17}$ crystallizes into the hexagonal structure with space group $P6_3$/mmc, where the magnetic $Nd^{3+}$ ions locate on the layered triangular lattice planes within the *ab*-plane and are stacked in an "AA"-type fashion along the *c*-axis. The low-temperature specific heat results show that no long-range magnetic ordering is detected down to 0.08 K.